\documentstyle[preprint,aps]{revtex}

\begin{document}

\draft
\preprint{RPI-N91-1994}
\title{ THE BLOOM-GILMAN DUALITY AND LEADING
LOGARITHMS$^*$}
\author{ Carl E. Carlson$^a$ and Nimai C. Mukhopadhyay$^b$\\
 ${}^a$ Physics Department, College of William and Mary\\
        Williamsburg, Virginia 23187\\
 ${}^b$ Physics Department, Rensselaer Polytechnic Institute\\
 Troy, NY 12180-3590\\}
\date{ July, 1994}

\maketitle

\begin{abstract}

The existing inclusive electroproduction data base
allows us a look at the issue of the relative behaviors of background and
resonance excitations,  a part of the Bloom-Gilman duality. These
data lack accuracy at high $Q^2$, but establish PQCD scaling
in the resonance region and even allow us a glimpse at
the leading logarithmic corrections due to the gluon radiation
and its possible quenching at large $W$ and $x$. These should inspire
better quality experimental tests at facilities like CEBAF II.
\end{abstract}



\section{Introduction }

Now that CEBAF is here, it is quite appropriate to ponder over the
question, {\em what if}  CEBAF had an electron beam of 8 GeV
energy and higher. Before such a prospect becomes  a realistic
proposal, this workshop seeks to identify interesting physics issues
that need {\em critically} such a higher energy facility.
Purpose of this paper is to point out one such high-$Q^2$ physics
problem that could be helped at such a facility: investigation of
the leading log effects from QCD in the resonance electroweak
 form factors\cite{ref1}. One of most fundamental issues in hadron
 physics is the possibility of reaching the perturbative domain
 of quantum chromodynamics(PQCD): in such a domain, the PQCD rules
 the behavior of the  electroweak structure function\cite{ref2}. There is still
 considerable debate on the domain of validity of the PQCD.

 \section{ The PQCD Rules for Helicity
Amplitudes for Electroexciting a Resonance}

 For the helicity amplitudes $G_{\pm ,0}$, defined in the Breit
 frame, for the electroproduction of a resonance, the PQCD rules
 tell us\cite{ref2}
 \begin{equation}
 G_i =\frac{g_i}{Q^n}
 \end{equation}
 where for $i=+, 0$ and $-$ respectively, $n=3$, $4$ and $5$, where
 $Q^2$ is the invariant mass squared and $g_i$'s are constants
 {\it modulo} $log Q^2$, dependent on the distribution amplitude of
 the relevant hadrons. In the amplitudes by Chernyak and
 Zhitnitsky, there is an accidental cancellation,
 making $g_+$ really small\cite{ref3}.
 Checking such a prediction is a fundamental enterprise
 in the resonance physics.
 Stoler's analysis\cite{ref4} leads support to this particular prospect
 for the Delta(1232), but this is not necessarily the case, accorrding to
   Davidson and Mukhopadhyay\cite{ref5} in a different
 approach,
 indicating intrinsic model dependence in this sort of analysis.

 Another interesting point in the PQCD  rule has been pointed out by the
  present authors\cite{ref6}. The above rules have basically assumed three
 quarks as the leading Fock components for the baryons. In the
 case of hybrids, which contain valence gluons, the transverse
 electroproduction is small, compared to ``normal" baryons. but the
 longitudinal one is not, and it scales like that for the normal
 baryons.  Similarly, characteristic high momentum transfer signatures
 of PQCD can tell whether hadrons such as
 the  $\Lambda (1405)$, $f_0(575)$ and $a_0(980)$  have
 normal three-quark (baryon), quark-antiquark(mesons) leading Fock
 configurations.

 \section{ The Bloom-Gilman Duality }
 First noted by Bloom and Gilman\cite{ref7}, more than two decades
 ago, this duality contains two parts:

(a) Resonances excited in the e-p scattering fall at roughly same
rate as the background underneath them with increasing $q^2=-Q^2$.

(b) The smooth scaling limit seen at high $Q^2$ and W for the
structure
function $\nu W_2(\omega^{\prime})$, where $\omega^{\prime}=1+W^2/
Q^2$, is an accurate  average for bumps seen at lower $Q^2$ and W,
but at same
$\omega^{\prime}$.

It is the second observation that connects the above behavior with
the classical definition of duality, originally proposed by
Dolan, Horn and Schmid\cite{ref8}.

The BG duality can be explained by the PQCD rules, as was shown by
DeR\'{u}jula, Georgi and Politzer\cite{ref9} and the present
authors\cite{ref2}. DeR\'{u}jula {\it et al.}
showed that the corrections to
the lower moments of the structure function due to final state
interactions
(or higher twist effects) are small, while the corrections to
higher moments are large. Thus, the average value of the structure
function cannot be very different from its values at high $Q^2$.
The present authors extended this concept to the longitudinal structure
function as well.

The BG duality has a strong significance for testing of the
PQCD rules in the resonance region: Due to the first part of the
duality, there is no need to separate the background and
resonance, as has been done by Stoler\cite{ref4}
and other authors\cite{ref10}  to test the PQCD rule. The structure
function {\it as a whole} can be checked for scaling.

Carlson and Mukhopadhyay\cite{ref1}  have examined the dual
relationship between the resonance and the scaling curves
for the inelastic structure function in the $Q^2$ range of 1 to 17
$(GeV/c)^2$ and found {\it no exception} to this rule.
Thus, $W_2$, averaged over a suitable range of the Bjoken $x$ (or the
equivalent Nachtmann $\xi$), yields the
smooth curve  seen at higher values of
$Q^2$ at the same value of $x$,  for  all resonance regions. {\it Looked
this way, even the Delta(1232), earlier suspected of ``misbehaving",
appears quite ``normal" or ``unexceptional".}

We have done the above analysis by dividing the resonance regions
into three zones, $1.12\leq W\leq 1.38$ the domain of the
Delta(1232); $1.38\leq W\leq 1.62$, the region containing
$N^*(1520)$ and $N^*(1535)$; $1.61\leq W\leq 1.80 GeV$, with
resonance bumps around 1.7 GeV. The analysis computes the
integrals
\begin{equation}
I_i=\int_{\Delta x_i}dx F_2 (x,Q^2),
\end{equation}
and \begin{equation}
S_i=\int_{\Delta x_i}dx F_2^{scaling}(x),
\end{equation}
where $\Delta x_i$ is the region of $x$ corresponding to the above
$W$ intervals. We have then computed the ratios $R_i=I_i/S_i$.
We have used the {\it constancy} of each $R_i$ as
the test for the BG duality. We
have also used the  Nachtmann variable $\xi =2x/(1+\sqrt{1+Q^2/\nu^2})$,
better for the inclusion of the target mass
effect and we  have tested  various scaling functions\cite{ref11}.
Our conclusion: {\em The BG duality  works nicely above $Q^2$ of the order of
four $GeV^2$,  in all three resonance regions!}

\section{ Leading Logs: Are Gluon Radiations Damped at High x?}
We have recently studied\cite{ref1} leading log corrections to the inelastic
scattering structure function at high Bjorken $x$. We first
investigate these corrections on the parton distribution function.
Starting with the Altarelli-Parisi equation having unsuppressed
gluon radiation, we find

\begin{equation}
q(x, t)=N_0(1-x)^{4+\frac{16}{3}(\ell n\ell n Q^2/\beta_1)} ,
\end{equation}
where q(x, t) is the quark distribution function of a given flavor ,
starting with the form
\begin{equation}
q(x, t_0)=N_0(1-x)^b,
\end{equation}
where $b$ is a constant, $t_0$ corresponds to some  benchmark $Q^2_0$;
symbols here have the usual meaning:
\begin{equation}
t=\ell n(Q^2/\Lambda^2),
\end{equation}
\begin{equation}
\beta_1=11-(\frac{2}{3}n_f)
\end{equation}
$n_f$, the number of fermion flavors,
\begin{equation}
\ell n\ell nQ^2=\ell n (\frac{\ell n(Q^2/\Lambda^2 )}{\ell n
(Q^2_0/\Lambda^2)})\equiv T(Q).
\end{equation}
We find, for $Q_0^2$ $=4$ $GeV^2$ and $Q^2\approx 20GeV^2$, $F_2$
changes by about 0.57 due to the logarithmic corrections in the Delta
region. Similar conclusions are reached for other resonances.

Moral: {\em Logs are important!}
\section{ The Bloom-Gilman Duality,
Leading Logs and Higher Twists}

Our observations can be summarized as follows:

(1) The inclusion of logarithmic effects helps to make the BG
duality idea work {\it better}.

(2). For distribution amplitudes due to Chernyak and Zhitnitsky  and King
and Sachrajda,
just to mention two we  have examined, the BG duality is logarithmically
violated.

(3). At $W>2GeV$ and high $x$ ($x>0.70$),
the {\em uncorrected} $(1-x)^3$ form
fits the data better, in agreement with an
argument due to Brodsky {\it et al.}\cite{ref12} that the
logarithms are {\em healed} in  the region where$(1-x)Q^2$ is small.

Points (1) and (3) will not contradict each other if we introduce a
$W$-dependent higher twist correction  such that we have
\begin{equation}
F_2\propto (1-x)^{3+\frac{16}{3}T(Q)/\beta_1}(1+C_2\frac{m_{N}^2}{W^2}).
\end{equation}
{}From the available data, we get
\begin{equation}
C_2=1.7.
\end{equation}

\section{ Concluding Remarks}

The existing inclusive electroproduction data in the resonance
region, poor though they are, still give valuable
insights into the Bloom-Gilman duality and the effects of
the gluonic radiation via logarithms. Log corrections seem to be
important in the resonance region, but at high $x$ and $W>2GeV$,
a plain $(1-x)^3$ fits the data. This requires a hypothesis
of evolution healing or the presence of higher twist effects. Measuring
structure function over a range of $x$ at fixed values of $W$ and
$Q^2$ respectively would deepen our insight into these mechanisms.

That brings us to CEBAF II. Amen to that!

\section{ Acknowledgement }
We are grateful for our  research support from the NSF(CEC)
and the U. S. Department of Energy (NCM). We thank S. Brodsky and
P. Stoler for many helpful discussions.

\vglue 0.3cm\noindent
$^*$  Invited talk at the 1994 CEBAF workshop,
presented by N. C. Mukhopadhyay.


\end{document}